# A self-stabilized coherent phonon source driven by optical forces

D. Navarro-Urrios[1,2]*, N. E. Capuj[3,4], J. Gomis-Bresco[1], F. Alzina[1], A. Pitanti[2], A. Griol[6], A. Martínez[5] and C. M. Sotomayor Torres[1,6]

[1]Catalan Institute of Nanoscience and Nanotechnology ICN2, Bellaterra (Barcelona), Spain

[2] NEST, Istituto Nanoscienze – CNR and Scuola Normale Superiore, Piazza San Silvestro 12, Pisa, I-56127

[3]Depto. Física, Universidad de la Laguna, La Laguna, Spain

[4] Instituto Universitario de Materiales y Nanotecnología, Universidad de La Laguna, La Laguna, Spain

[5]Nanophotonics Technology Center, Universitat Politècnica de València, Spain

[6]Catalan Institute for Research and Advances Studies ICREA, Barcelona, Spain

*Correspondence to: daniel.navarrourrios@nano.cnr.it

Optical forces can set tiny objects into states of coherent mechanical oscillation, also known as mechanical or phonon lasing. We present a novel pumping mechanism in an opto-mechanical photonic crystal that realizes mechanical lasing with relaxed requirements for the optical-mechanical modes and their inter-coupling. It derives from a spontaneously triggered thermal/free carrier self-pulsing and the anharmonic modulation of the radiation pressure force that comes as a consequence. Moreover, the feedback of the mechanics on the self-pulsing frequency-entrains both oscillators, creating a self-stabilized indecomposable system. A manifold of frequency-entrained regions with two different mechanical modes (at 54 and 122 MHz) are observed as a result of the wide tuneability of the natural frequency of the self-pulsing. The system operates at ambient conditions of pressure and temperature in a silicon compatible platform, which enables its exploitation in sensing, intra-chip metrology or time-keeping applications.

Coherent creation, manipulation and detection are key ingredients for the full technological exploitation of particle excitations. While lasers and masers have dominated among the sources in electromagnetism, the natural lack of discrete phonon transitions in solids make the realization of coherent vibration sources a formidable challenge. The introduction of miniaturized self-sustained coherent phonon sources is crucial in applications such as mass-force sensing *(1)* and intra-chip metrology and time-keeping *(2)*. They also offer intriguing opportunities for studying and exploiting synchronization phenomena among several oscillators by introducing a weak interlink *(3,4)*, which can be optical *(5,6)* or mechanical *(7)*.

In recent years several versions of "phonon lasers" have been reported in electro- and opto-mechanical systems *(1,8-10)*. The most common optical pumping mechanism is by means of the retarded radiation pressure within a nano/microcavity through a sideband-assisted energy scattering process *(11)*. The stringent requirement to reach the lasing threshold, e.g. unitary cooperativity *(12)*, restricts its effective operation to high quality factor modes with large opto-mechanical (OM) coupling strengths. Alternatively, an external feedback can be used to parametrically increase the mechanical oscillations up to the lasing regime, where the technological requirements of electronic control limits the maximum mechanical frequency operation *(13, 14)*.

In this Letter, we report an integrated coherent phonon source using a completely different injection scheme that allows for lasing with cooperativity values as low as $10^{-2}$. It extracts energy from a cw infrared laser source and is based on the triggering of a thermo-optical/free-carrier-dispersion self-pulsing limit-cycle (*15-18*), which modulates the radiation pressure forces. The large amplitude of the coherent mechanical motion acts as a feedback that stabilizes and entrains the self-pulsing oscillations to simple fractions of the mechanical frequency.

The device presented here (*19*) is a one dimensional OM photonic crystal fabricated using standard silicon (Si) nanofabrication tools (see Supplementary Discussion 1). As seen from the SEM top-view of Fig. 1A, the crystal lattice constant is quadratically reduced towards the center of the beam, hereby defining optical defect modes.

Low-power, optical transmission measurements of the second order even-even optical mode (see Finite-Element-Method (FEM) simulation in inset of Fig. 1B), show a Lorentzian lineshape with a cavity resonant wavelength at $\lambda_r$ = $\lambda_o$ = 1527.01 nm and a linewidth $\Delta\lambda_o$ =70 pm, holding an optical Q-factor $Q_o$= 2.2×$10^4$ (Fig. 1B). The same measurement at high input powers shows the typical bistability "saw-tooth" shaped transmission, caused by the red-shift of the resonance frequency due to thermo-optic (TO) nonlinearities *(20)*. In this regime, decreasing the laser-cavity detuning from the blue-side corresponds to an increase of the intracavity photon number ($n_o$).

Thermally driven motion of the low frequency mechanical modes is seen by processing the transmitted light with a spectrum analyzer. Here, mechanical modes with a non-negligible optomechanical coupling rate ($g_{OM}$) appear as narrow Lorentzian peaks in the frequency spectrum, as reported in Fig. 1 (c). The odd, in-plane flexural modes of the beam have a single particle coupling rate ($g_{o,OM}$) of tens of KHz, as obtained from FEM simulations in a perturbative approach for shifting material boundaries (*21*). The even-mode family is weakly coupled due to symmetry consideration. In this work, we focus on the second (three anti-nodes) and third (five anti-nodes) order odd modes, which have a center frequency of $\nu_{m,2}$=54 and $\nu_{m,3}$=122 MHz, simulated effective masses $m_{eff,2}$=2.4x$10^{-15}$ kg and $m_{eff,3}$=2.6 x$10^{-15}$ kg, and mechanical Q-factors $Q_{m,2}$=450 and $Q_{m,3}$=520, respectively. The Q-factors are mainly limited by the experimental conditions: room-temperature (thermo-elastic damping) and atmospheric pressure.

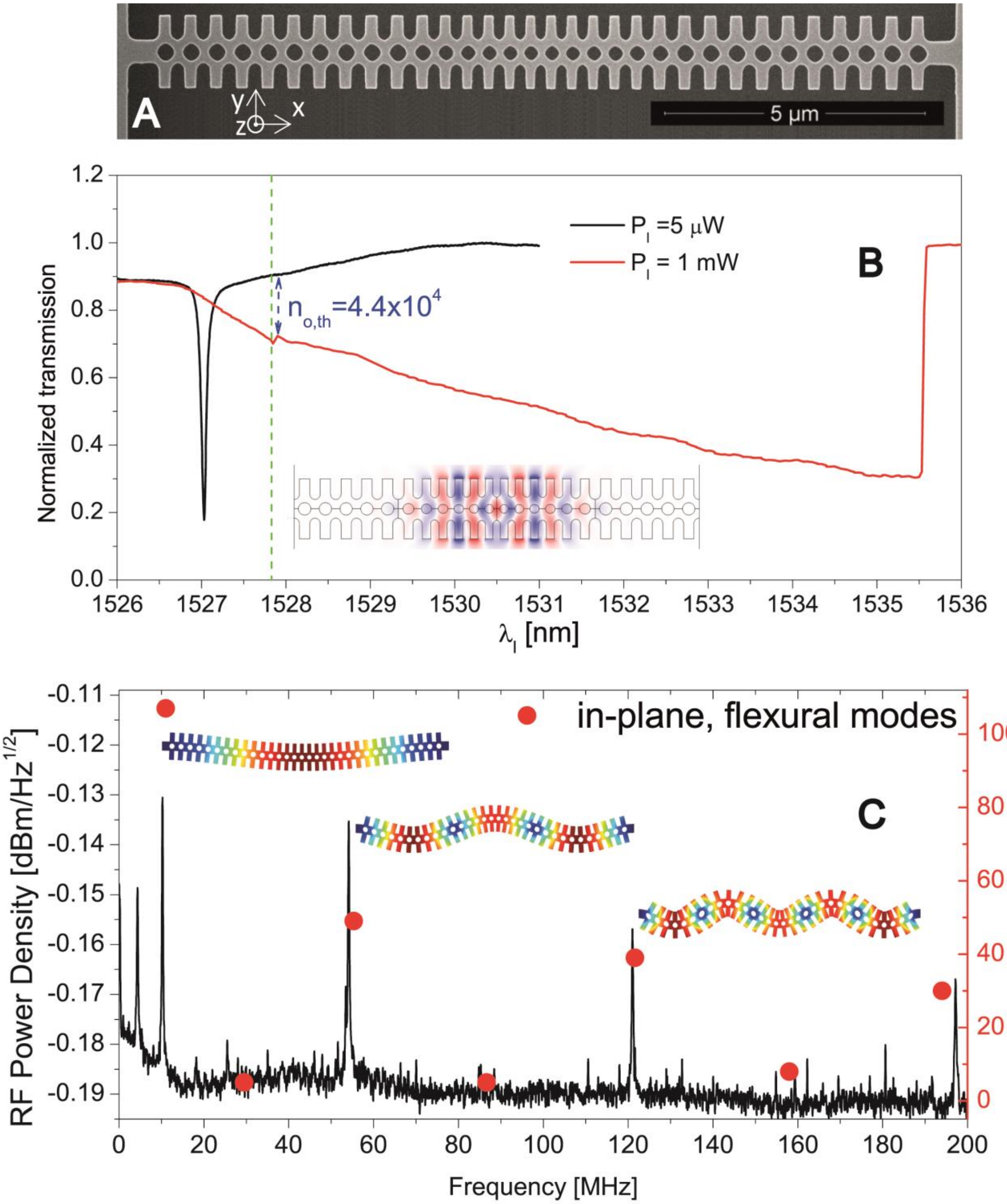

**Fig. 1. Optical and mechanical responses of the OM photonic crystal. (A)** SEM micrograph of the OM photonic crystal. **(B)** Normalized transmission spectrum around the studied optical

mode for input laser powers of 5 μW and 1 mW (black and red curves respectively). The threshold for establishing the regime of self-sustained oscillation is indicated (green dashed vertical line). **(C)** Transduced acoustic modes up to 200 MHz. The red dots correspond to the $g_{o,OM}/2\pi$ values of the in-plane flexural eigenmodes as predicted by FEM simulations. The OM photonic crystal deformation profiles associated to the first three odd in-plane flexural modes are also illustrated.

Temporally resolving the optical transmission with a fast detector reveals a complicated nonlinear dynamics when the intracavity photon number is above $n_o=n_{o,th}=4.4\text{x}10^4$ (black curve of Fig. 2A). It is well known that the dynamic competition among different nonlinearities can lead to instabilities, self-sustained oscillations and chaotic behavior of light within a photonic device *(22-24)*. The two main sources of nonlinearities in Si resonators are Free-Carrier-Dispersion (FCD) and TO effect. In the former, the excess of free-carriers leads to a reduction of the material refractive index and therefore a blue-shift of the optical cavity *(25)*. On the other hand, the TO relates an increase of the refractive index with an increase of the material temperature. Since the main source of heating is due to the free-carrier absorption, the dynamics of Free-Carrier Density (*N)* and the Temperature Increase (Δ*T)* are linked together and can be described by a system of coupled rate equations (*15,16*):

$$\begin{aligned} \frac{dN}{dt} &= -\frac{1}{\tau_{FC}}N + \beta\left(\frac{hc^3}{n^2\lambda_o V_o^2}\right)n_o^2; \\ \frac{d\Delta T}{dt} &= -\frac{1}{\tau_T}\Delta T + \alpha_{FC} N n_o; \end{aligned} \qquad (1)$$

, where $h$ is the Planck constant, $c$ is the speed of light and $V_o$ is the optical mode volume. In the FCD equation (first row) we consider a Two-Photon Absorption (TPA) generation term (*16*), where $\beta$ is the tabulated TPA coefficient *(20)* and a surface recombination term governed by a characteristic lifetime $\tau_{FC}$. The TO equation (second row) reflects the balance between the fraction of photons that are absorbed and transformed into heat due to free-carrier-absorption ($\alpha_{FC}$ is defined as the rate of temperature increase per photon and unit free-carrier density) and the heat dissipated to the surroundings of the cavity volume, which is governed by a characteristic lifetime $\tau_T$. All is rescaled through the intracavity photon number $n_o$, which includes the cavity filtering effect:

$$n_o = \left(2\,P_l\frac{\kappa_e}{\kappa^2}\frac{\lambda_o}{hc}\right)\frac{\Delta\lambda_o^2}{4\left(\lambda_l-\lambda_r\right)^2+\Delta\lambda_o^2} \quad (2),$$

where the term in parentheses represents the maximum $n_o$ achieved during the oscillation, $\kappa_e$ and $\kappa$ are the extrinsic and total optical decay rates respectively, and $\lambda_l$ and $P_l$ are the laser wavelength and input power respectively *(26)*. Importantly, $\lambda_r \approx \lambda_o - \frac{\partial\lambda_r}{\partial N}N + \frac{\partial\lambda_r}{\partial T}\Delta T$ is the cavity resonant wavelength including first order nonlinear effects. Note that in Eq. (1) we have neglected spatial temperature gradients ($\partial\Delta T/\partial x$, $\partial\Delta T/\partial x=0$) and assumed a homogenous distribution of $N$. Moreover, we consider an adiabatic response of the optical mode to the refractive index change: this comes naturally when we take into account the very different time-scale of the FCD and TO effects with respect to the optical radiative decay of ~20 ps.

For particular sets of conditions, $N$ and $\Delta T$ form a stable limit cycle (Supplementary Discussion 4), known as self-pulsing (SP) *(15)*. This phenomenon has been observed in various photonic devices, such as microdisks and photonic crystals *(15-18)*. In our experiment, when $n_o$ is above $n_{o,th}$, the system enters in the SP regime and the transmission assumes an asymmetric, double dip shape coming from the interplay between fast (FCD) and slow (TO) dynamics. Numerical integration of Eq. 1 reproduces very well the experimental data, as reported in Fig. 2A. The characteristic lifetimes giving the best agreement with the experimental data are: $\tau_T$~0.5 μs and $\tau_{FC}$~0.5 ns, which are compatible with values reported elsewhere (*16*). Interesting insights can rise from the inspection of the $\frac{\partial \lambda_r}{\partial N} N$ (FCD) and $\frac{\partial \lambda_r}{\partial \Delta T} \Delta T$ (TO) simulated time traces of Fig. 2C. Within the limit cycle the temperature is prevented to reach the steady state, which reflects in the triangular waveform of Fig. 2C (red curve). The slow temperature decay (mirrored to a wavelength increase in the panel) dominates $\nu_{SP}$, being the Free-Carrier creation and recombination very fast at these time scales. Since $\Delta T$ instantaneous decay rate depends on the temperature absolute value, $\nu_{SP}$ can be enhanced up to five times by increasing the amount of total heat in the cavity. In our system this is simply done by increasing $\lambda_l$, which increases the time-averaged intracavity photon number. In order to increase further its operation regime, a reasonable strategy is to speed up the thermal/free-carrier dynamics by increasing the thermal diffusivity of the OM structures and the free-carrier recombination rate.

When the self-pulsing is active, the light within the cavity is modulated in a strongly anharmonic way, creating an "optical comb" in the frequency domain with multiple peaks spectrally located at integers of the SP oscillating frequency $\nu_{SP}$ (Supplementary Discussion 5). This translates in a modulation of the radiation pressure optical force, which can be easily evaluated as a function of the intracavity photon number $F_o = \hbar g_{OM} n_o$. The flexural modes of the nanobeam can be described as damped linear harmonic oscillators driven by the anhamonic force:

$$m_{eff} \frac{d^2 u}{dt^2} + m_{eff} \frac{2\pi \upsilon_m}{Q_m} \frac{du}{dt} + k_{eff} u = F_o, \quad (3)$$

where $u$ is the generalized coordinate for the displacement of the mechanical mode and $k_{eff}$ is its effective spring constant. Finally, the nonlinear wavelength has to include now the effect of the mechanical motion when evaluating Eq. 2, as:

$$\lambda_r \approx \lambda_o - \frac{\partial \lambda_r}{\partial N} N + \frac{\partial \lambda_r}{\partial T} \Delta T + \frac{\lambda_o^2 g_{OM}}{2\pi c} u \quad (4)$$

Importantly, the response of $n_o$ to deformation is also adiabatic since the radiative lifetime of the optical mode (~ 20 ps) is three orders of magnitude smaller than the mechanical oscillation period (~ 20 ns).

The dynamics of the two systems described by Eqs. (1) and (3) are coupled through the number of photons in the cavity. The most significant feature of the coupled system is that self-sustained mechanical motion is achieved if one of the low harmonics of the SP main peak at $\nu_{SP}$ is resonant with the mechanics ($M\nu_{SP} = \nu_m$, where $M$ is an integer number) (Supplementary Discussion 5). In fact, the coherence of the mechanical oscillation is maintained since the mechanical mode lifetime (~10 μs) is much larger than $1/M\nu_{SP}$ (~ 20 ns). As an example, Fig. 2B shows experimental and theoretical data (black and red curves respectively) for the case corresponding to M=2. As expected, the mechanical oscillation at frequency $\nu_m$ (vertical lines) is overimposed on a SP trace at frequency $\nu_m/2$. Separating all the terms contributing to $\lambda_r$ (Fig. 2D), we see that a simple harmonic signal (blue curve denoted by OM) is included with the FCD and TO signal to obtain the data of Fig. 2B. Our model reproduces quite well the dynamics of the transmitted signal, revealing that the contribution to $\lambda_r$ of the displacement $u$ is of similar magnitude of the ones from $N$ and $\Delta T$. The coherent oscillations of the mechanics implies that the driving strength overcomes the mechanical dissipation in the system, entering a regime that has been defined as "phonon lasing"*(2)*, although the particle generation mechanism is very different to the one in photon systems. In this regime, phonons are created coherently, as seen from the self-sustained oscillations of Fig. 2B. It is worth noting that, although the same phonon lasing regime can be achieved through dynamical back-action processes, the injection mechanism reported here is intrinsically different and relaxes the strict requirement needed to enter this state (unitary field-enhanced cooperativity, defined as $C = n_o \frac{4g_{o,OM}^2}{\kappa} \frac{Q_m}{2\pi\upsilon_m}$) *(12)*. In fact, using the SP pumping mechanism, mechanical lasing is achieved in this work with $C$ values within the $10^{-2}$ range.

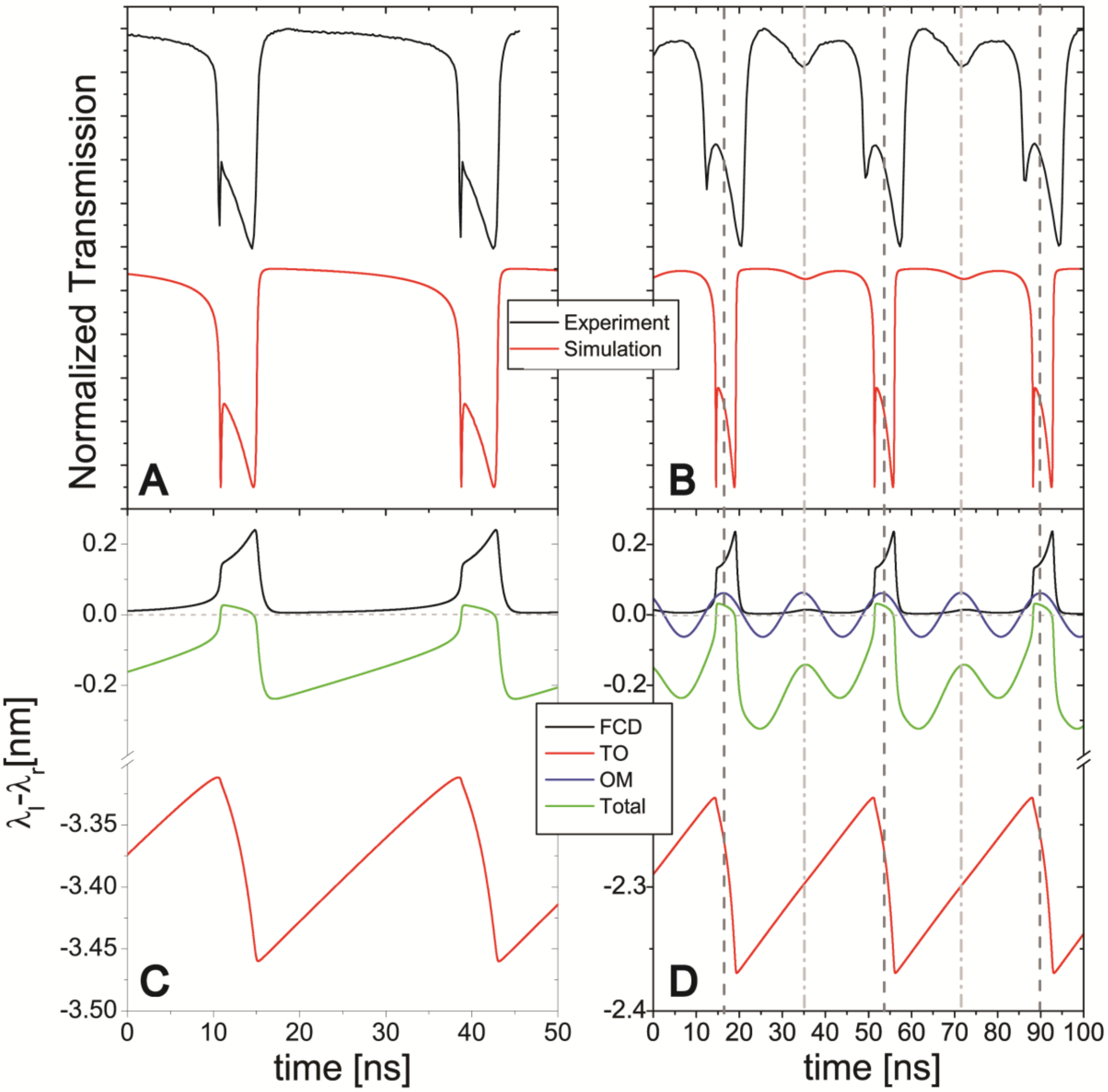

**Fig. 2. Dynamics of the coupled system for $P_{in}$=1 mW and $n_o$ over the threshold. (A-B)** The black (red) curve shows the experimental (simulated) temporal profile of the transmitted signal obtained at $\lambda_l$ = 1530.2 nm (panel A) and at $\lambda_l$ = 1529.1 nm (panel B). In the latter case the OM photonic crystal is oscillating coherently in its 3$^{rd}$ in-plane flexural mode. The SP is frequency-entrained with the mechanical oscillation (M=2) **(C-D)** Simulated temporal profiles of FCD (black), TO (red) and OM (blue, only in panel D) contributions to the spectral shift of $\lambda_r$ obtained at $\lambda_l$ = 1530.2 nm (panel C) and at $\lambda_l$ = 1529.1 nm (panel D). The overall spectral shift of $\lambda_r$ with respect to $\lambda_l$ is represented in green. The dashed horizontal line indicates the resonant condition, i.e., $\lambda_l = \lambda_r$.

The full RF power spectrum as a function of $\lambda_l$ presents several regions and features of interest. As shown in Fig. 3A, below threshold (left of the dashed vertical green line) only peaks related to the transduction of the mechanical modes appear. Above threshold, the signature of SP emerges. When the mechanics/self-pulsing resonant condition is verified, "flat regions" appear, indicating the coherent vibration of the OM photonic crystal. Since all the dynamics is coupled together, the OM oscillations provide an active feedback that stabilizes the SP. In those specific conditions, the two oscillators are frequency-entrained (FE) in a way that the SP adapts its oscillating frequency to be a simple fraction of the mechanical eigenfrequency. Similarly to the case of synchronized oscillators, the lowest M values have the largest FL zones (*3*). Inspecting the spectra extracted along the white vertical lines of Fig. 3A and reported in Fig. 3B, it is clear that the linewidth in the FE regions (III - black curve) is orders of magnitude narrower than both the SP oscillation (II - red curve) and the mechanical thermal Brownian motion (I - green curve), as expected from a coherent source. Following the SP main frequency peak evolution (M=1), we can extract a curve that resembles the Devil's staircase of synchronized systems *(3)*, as reported in Fig. 3C. Several regions of FE appear both with the 2[nd] and 3[rd] in-plane flexural odd modes (M=1, 2, ..., 6 and M=3, 4 respectively), where $\nu_{SP}$ is stable (see Supplementary Videos 1 and 2) and robust to wide variations of $\lambda_l$ and/or the laser power. By tuning $\lambda_l$ it is possible to sweep over different FE states and switch to frequency-unlocked situations.

These experimental features can be well reproduced (green curve of Fig. 3C) by introducing a $g_{OM}$ in good agreement with the evaluated one reported in Fig. 1C. As expected, if the active feedback is turned off (i.e., $g_{OM}=0$), all the FE regions disappear and the SP naturally evolves with no interaction with the mechanics. The differences between model and experiment with regard to the spectral width of plateaus with M>1 is associated to the non-linearity of the elastic restoring force and the damping term *(27)*.

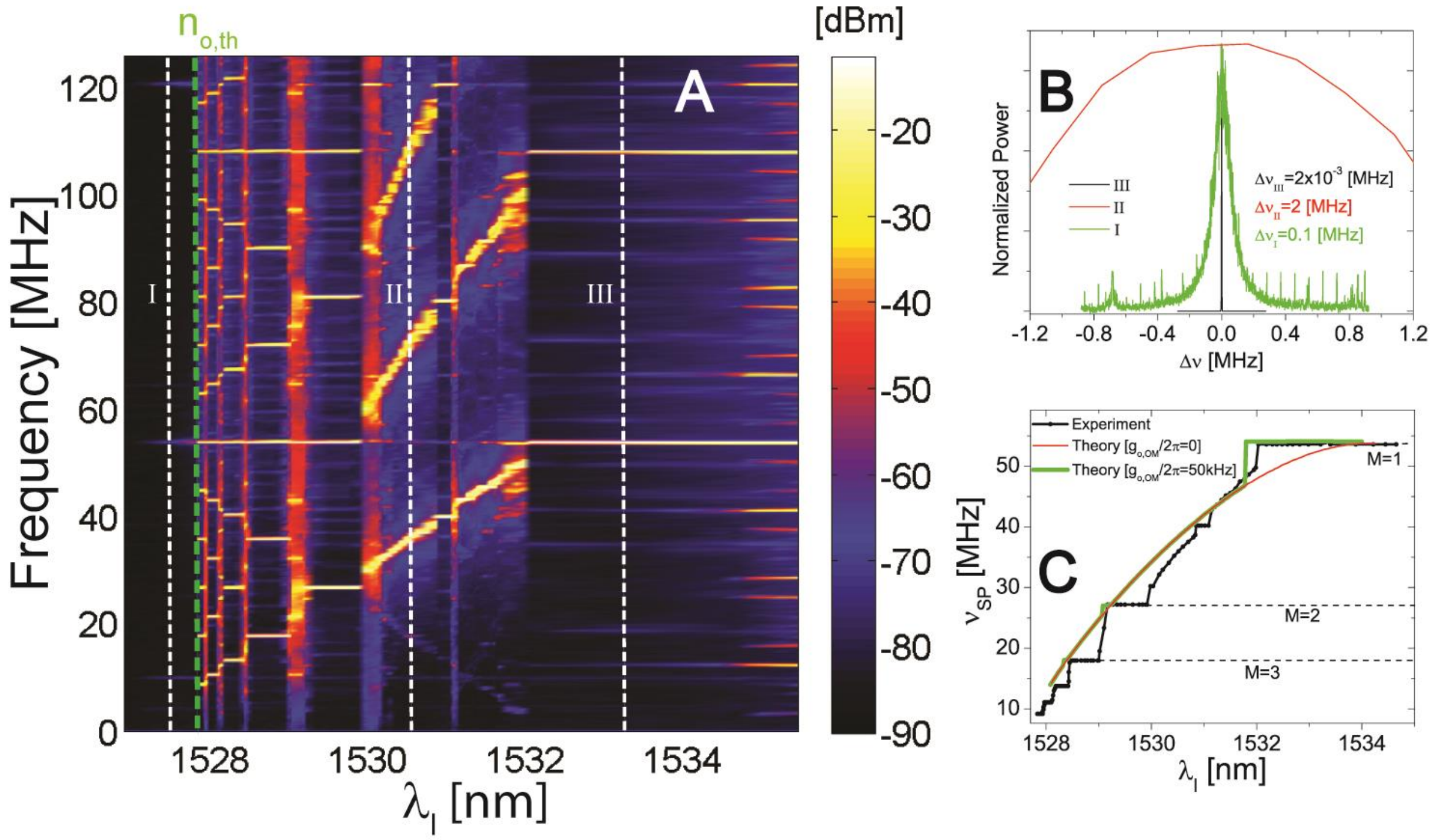

**Fig. 3. Multiple frequency entrainment and phonon lasing. (A)** Color contour plot of the RF power obtained for a $P_l$=1 mW. The green-dashed line indicates the threshold position. The vertical white-dashed lines indicate situations in which the signal corresponds to: thermally

activated motion below threshold (I), SP within a frequency-unlocked regime (II) and frequency-entrained situation for M=1 (III). **(B)** Normalized RF spectra corresponding to the cuts I, II and III of panel A (green, red and black curves respectively), where I illustrates the 2nd in-plane flexural odd mode and II and III the first harmonic of the transduced signal. **(C)** Simulated behavior of $\nu_{SP}$ considering only the 2nd mode and its estimated value of $g_{o,OM}$ (green line). The red curve is the expected SP behavior in the absence of feedback, i.e., if the OM photonic crystal were not free to oscillate mechanically or uncoupled with the electromagnetic field. The experimental curve is also included (black dotted line).

The mechanism of generation of coherent mechanical oscillations showed in this work, with its self-stabilized nature, could represent a viable solution to generate and feed phonons to waveguides or membranes. It provides a platform to perform fundamental experiments using phonons, such as the analogous to the Young's double-slit experiment. Since it operates on a silicon compatible platform at atmospheric conditions it can represent a first building block in the implementation of integrated phononic circuits. Among other functionalities, these mechanical oscillators can be used as ‘clocks’ to provide reference signals. In this context, synchronization phenomena among networks of self-sustained mechanical oscillators could be exploited to extend further the spatial range of a common coherent signal.

**Acknowledgments:** This work was supported by the EC through the project TAILPHOX (ICT-FP7-233883) and the ERC Advanced Grant SOULMAN (ERC-FP7-321122) and the Spanish project TAPHOR (MAT2012-31392). The authors sincerely thank B. Djafari-Rouhani, Y. Pennec and M. Oudich for the design of the OM photonic crystal, and A. Trifonova and S. Valenzuela for a critical reading of the manuscript. A. M and A. G thank L. Bellieres and N. Sánchez-Losilla for their contributions in the OM photonic crystal etching processes.

**Supplementary Materials:**

Materials and Methods; Figures S1-S11; References (S*1*-*S6*); Movies S1-S3

## Materials and methods

### S1. Devices

The investigated device is an optomechanical (OM) photonic crystal whose unit-cell contains a hole in the middle and two symmetric stubs on the sides (Fig. S1). The peculiarity of this geometry resides in having a full phononic band-gap at ~4GHz (*1*). The investigation of high frequency mechanical modes is reported elsewhere *(2)*.

For the sake of clarity, we report here the geometrical parameters of our device. In a defect region consisting of 12 central cells the pitch (a), the radius of the hole (r) and the stubs length (d) are decreased in a quadratic way towards the center. The maximum reduction of the parameters is denoted by Γ. At both sides of the defect region a 10 period mirror is included. The nominal geometrical values of the cells of the mirror are a=500nm, r=150nm, and d=250nm. The total number of cells is 32 and the whole device length is about 15μm. Fig. S1 shows a SEM micrograph of one of the OM photonic crystals.

We have fabricated a set of devices in which Γ has been varied from Γ=64% to Γ=83% of the original values. All the results presented in this work correspond to the structure with Γ=83%, but the same effects have been observed for other values of Γ.

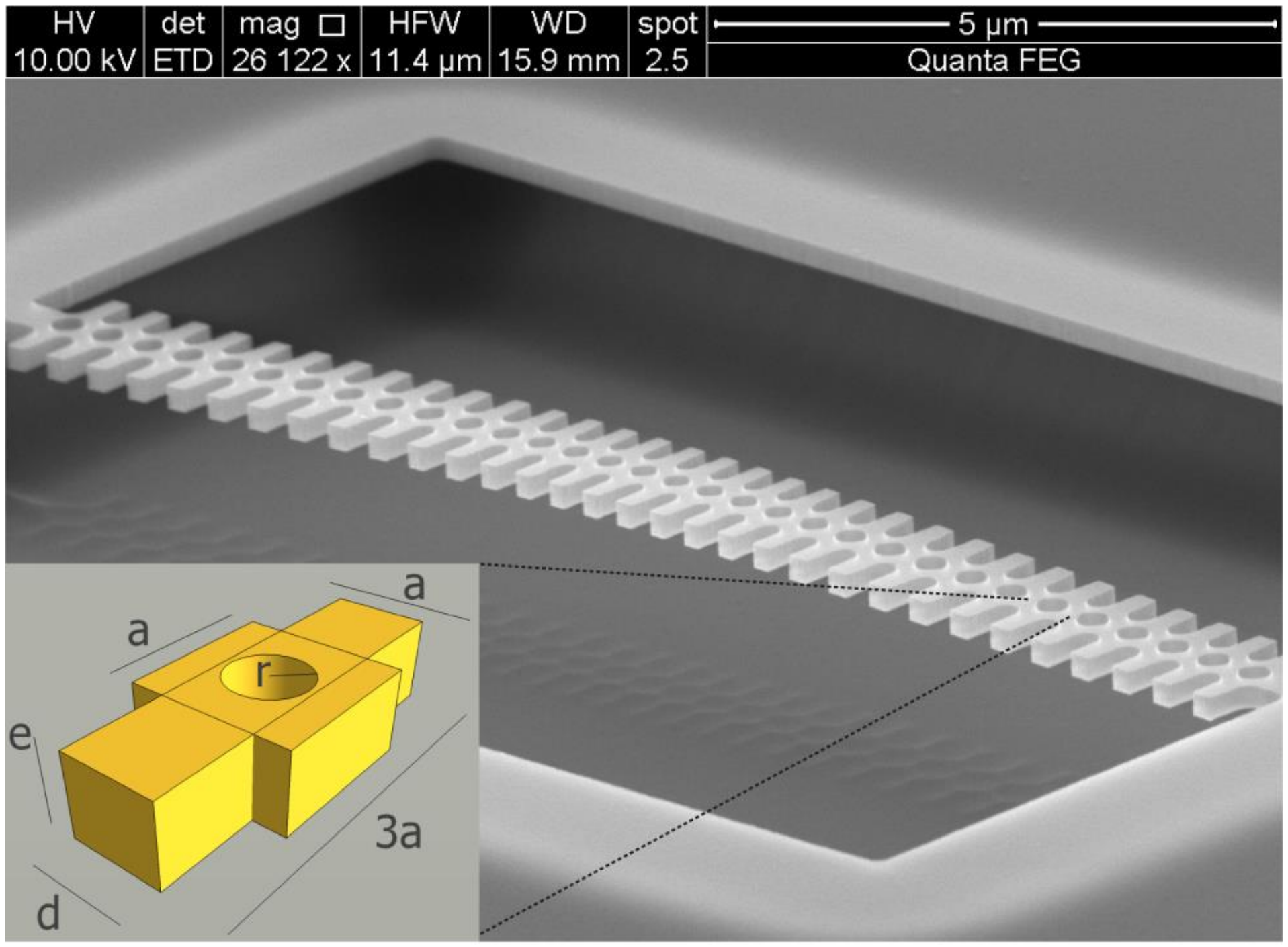

**Fig. S1.** SEM micrograph of a Γ=83% device. A sketch of the unit-cell is included as an inset.

The devices were fabricated in standard silicon-on-insulator (SOI) SOITEC wafers with silicon layer thickness of 220 nm (resistivity ρ ~1-10 Ω $cm^{-1}$, p-doping of ~$10^{15}$ $cm^{-3}$) and a buried oxide layer thickness of 2 μm. An electron beam direct writing is done on a 100 nm Poly-methyl-methacrylate (PMMA) resist film. After pattern transfer into silicon by Reactive Ion Etching (RIE) process, a BHF silicon oxide etch is carried out to remove the buried oxide layer and release the beam structures.

## S2. Optomechanical coupling calculation

The calculation of $g_{o,OM}$ is done using the integral given by Johnson et al. (2) for the moving interfaces effects:

$$g_{o,OM} = -\frac{\pi\lambda_o}{c}\frac{\oint (\mathbf{Q}\cdot\hat{\mathbf{n}})\left(\Delta\varepsilon E_{||}^2 - \Delta\varepsilon^{-1} D_{\perp}^2\right) dS}{\int E\cdot D dV}\sqrt{h/2m_{eff}\nu_m} \quad \text{(S1)}$$

Where **Q** is the normalized displacement (max{|**Q**|}=1), $\hat{\mathbf{n}}$ is the normal at the boundary (pointing out), **E** is the electric field and **D** the electric displacement field. $\varepsilon$ is the dielectric permittivity, $\Delta\varepsilon=\varepsilon_{silicon}-\varepsilon_{air}$ , $\Delta\varepsilon^{-1}=\varepsilon^{-1}_{silicon}-\varepsilon^{-1}_{air}$. $\lambda_o$ is the cold optical resonance wavelength, c is the speed of light in vacuum, h is the Planck constant, $m_{eff}$ is the effective mass of the mechanical mode and $\nu_m$ is the mechanical mode eigenfrequency.

## S3. Experimental set-up

The experiments are performed in a standard set-up for characterizing optical and mechanical properties of OM devices.

A tunable infrared laser covering the spectral range between 1460-1580nm is connected to a tapered fiber. The polarization state of the light entering the tapered region is set with a polarization controller. The thinnest part of the tapered fiber is placed parallel to the OM photonic crystal, in contact with an edge of the etched frame (top right photo of Fig. S2a, Fig. S2b). The gap between the fiber and the structure is about 200 nm, as roughly extracted from geometrical considerations using the radius of the fiber loop and the contact point position. A polarization analyzer is placed after the tapered fiber region.

The long tail of the evanescent field and the relatively high spatial resolution (~5 $\mu m^2$) of the tapered fiber allows for the local excitation of resonant optical modes of the OM photonic crystal. Once in resonance, the mechanical motion activated by the thermal Langevin force causes the transmitted intensity to be modulated around the static value (Fig. S2c).

To check for the presence of a radiofrequency (RF) modulation of the transmitted signal we have used an InGaAs fast photoreceiver with a bandwidth of 12 GHz. The RF voltage is connected to the 50 Ohm input impedance of a signal analyzer with a bandwidth of 13.5 GHz.

All the measurements have been performed in an anti-vibration cage at atmospheric conditions of air pressure and temperature.

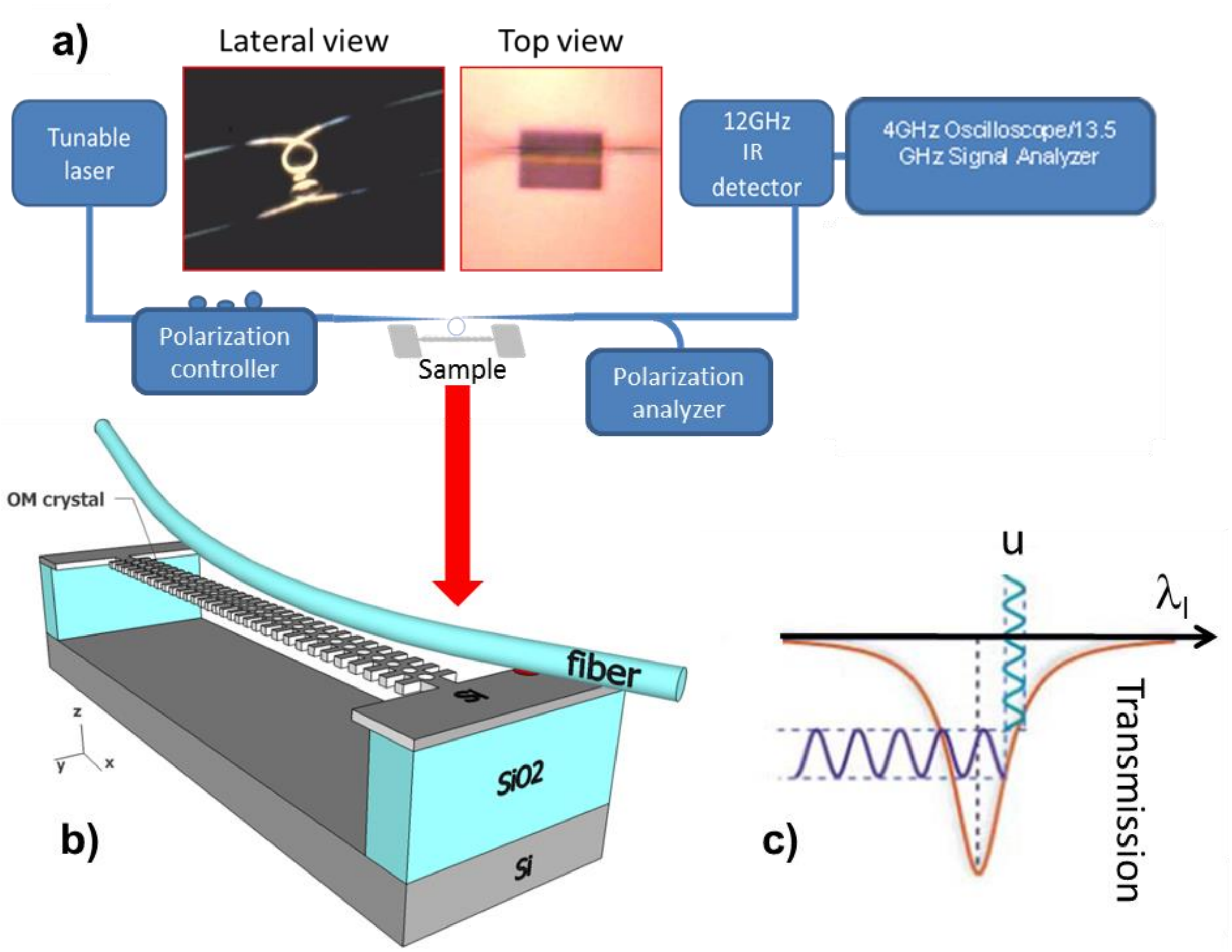

**Fig. S2. a)**. Sketch of the experimental setup to measure the optical and mechanical properties of the OM devices. The sample size has been greatly increased for clarity. The top left photo (labelled "Lateral view") shows a lateral view of the real microlooped tapered fiber close to the sample, where the fiber can be seen reflected on the sample. The top right photo (labelled "Top view") shows a top view of the tapered fiber placed parallel with the OM structure and in contact with one of the edges of the etched frame. **b)**. Relative positioning of the tapered fiber and the OM photonic crystal. The leaning point of the fiber is highlighted in red. The fiber is placed close enough to the central part of the OM photonic crystal to excite efficiently its localized photonic modes. **c)** Scheme of the transduction principle.

### *S3.1. Tapered fiber characteristics and fabrication procedure*

The experiments are carried out with tapered optical fibers having diameters in the smallest section of about 1.8μm (Fig. S3a), which is commensurate with the wavelength of interest (around 1.5 μm) and ensures an evanescent field tail of several hundreds of nanometers.
For the fiber fabrication, we used a home-made setup in which a SMF-28 optical fiber is stretched in a controlled way using two motorized stages. The central part of the fiber is placed in a microheater where the temperature is about 1180$^{o}$C *(2)*.
The fiber transmission at a wavelength of 1.5 μm is monitored during the pulling procedure (Fig. S3b). The signal is subjected to a short time Fast Fourier Transform(FFT) algorithm, so that the frequency components associated to inference between different supported modes are measured (Fig. S3c). The single mode configuration is achieved when all those frequency components disappear (Fig. S3d).

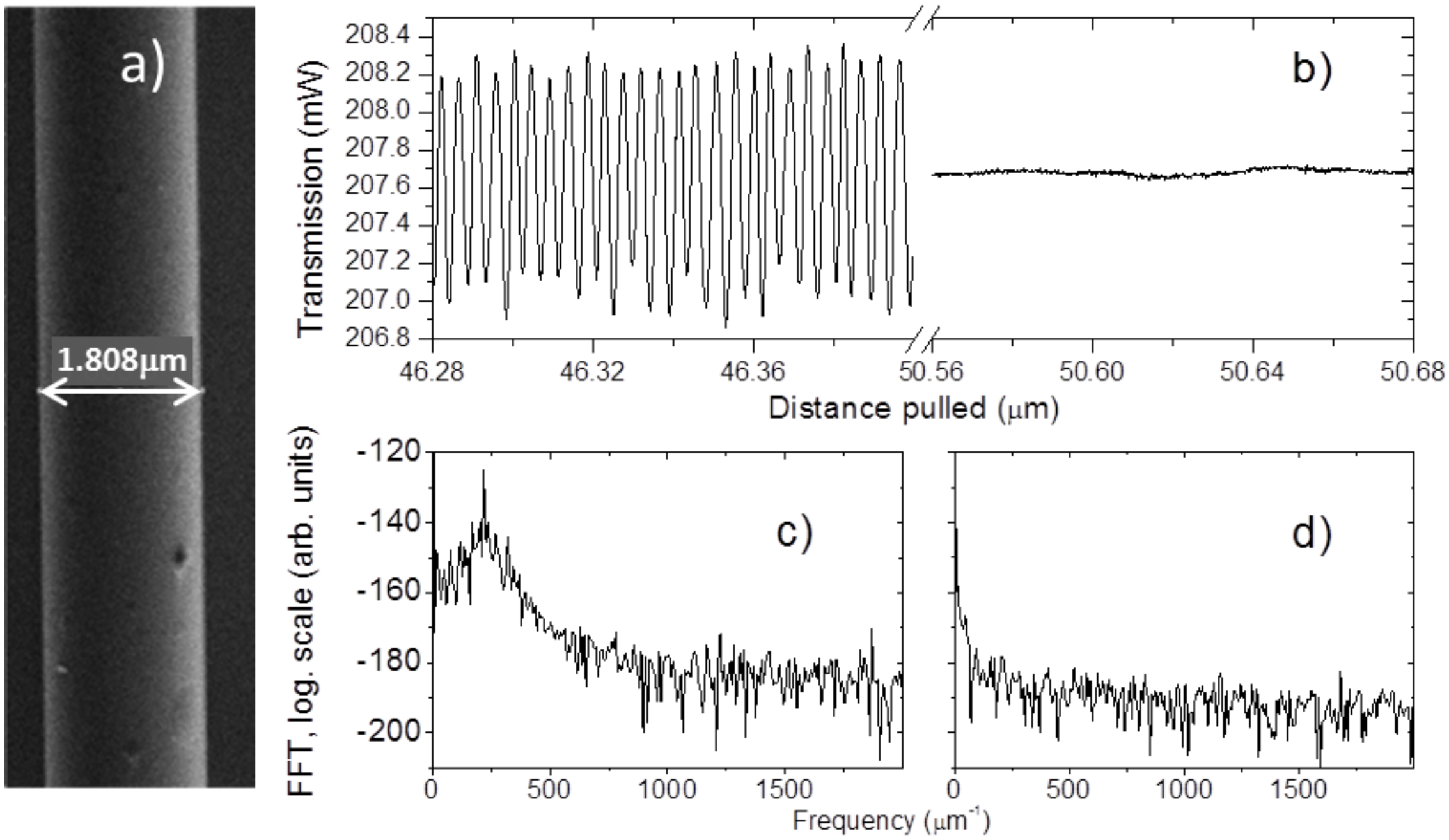

**Fig. S3. a)** SEM image of the thinnest part of a tapered fiber. **b)** Transition from multimode to monomode while pulling the two extremes of the fiber as seen in temporal scale. **c)** FFT of the transmitted signal before the transition to single mode. **d)** FFT of the transmitted signal after the transition to single mode.

Using two rotating fiber clamps, the tapered fiber is twisted twice around itself. The two fiber ends are gently brought closer over several hundreds of micrometers so that a looped structure forms in the tapered region. The two fiber ends are afterwards pulled apart to reduce the loop size down to a few tens of µm. In that process, the two parts of the fiber at the loop closing point (upper part of the loop on the top left photo Fig. S2) slide smoothly in opposite senses. The micro-looped shape provides functionalities similar to those of dimpled fibers (*5*).
For the fiber loop photonic structure, the dispersion relation is linear and the group refractive index ($n_g$) is equivalent to the effective refractive index ($n_{eff}$). We have calculated $n_{eff}$ using the Beam Propagation Method taking the material refractive index of the cladding of the initial fiber to be n=1.468 at $\lambda_l$=1515nm, resulting in a value of $n_{eff}$=1.373.

## S4. TO/FCD self-pulsing

### *S4.1. General behavior.*

We have implemented the model first reported by Johnson et al. (6), which can describe the dynamics of the self-pulsing (SP) mechanism (Eq. 1 of the main text).

The fitting parameters used to reproduce the frequency-unlocked cases are $\tau_T$=0.5 [μs], $\tau_{FC}$=0.5 [ns] and $\alpha_{FC}$=4x10$^{-13}$ [K cm$^3$ s$^{-1}$], while the initial conditions verify that $\frac{\partial \lambda_r}{\partial \Delta T}\Delta T(0)=\lambda_l-\lambda_o$. Thermo-optic (TO) and Free-Carrier-Dispersion (FCD) coefficients were independently calculated by assuming that the observed wavelength shift is only associated to an average change in the Si refractive index within the region overlapped with the electromagnetic fields and using tabulated values for its dependence with temperature and free-carrier density. This procedure lead to the following values: $\frac{\partial \lambda_r}{\partial N}$=7x10$^{-19}$ [nm cm$^3$] and $\frac{\partial \lambda_r}{\partial \Delta T}$=6x10$^{-2}$ [nm K$^{-1}$].
Hereafter, we compare the dynamics of the optical transmission, $N$, $\Delta T$ and $\Delta\lambda$ in the same time scale (Fig. S4a, panels A, B, C and D respectively). These results are equivalent to those shown in Fig.2C. The SP dynamics is illustrated by the temporal evolution of the resonance-shift ($\lambda_l$-$\lambda_r$), which contains the overall result of adding the different contributions defining $\lambda_r$, i.e., $\lambda_o$, FCD and TO effects. An anharmonic behaviour can be observed, with a main oscillation frequency at $\nu_{SP}$. Four specific points of the temporal curves of Fig. S4 are marked with orange numbers:

- On **point 1**, a fast increase of $N$ starts due to the absorption of part of the optical energy stored in the cavity.
- The first minimum observed in transmission occurring at **point 2** is related to the quick part of the increase of $N$, which leads to a sign change of $\lambda_l$-$\lambda_r$ from negative to positive values ($\lambda_r$ blue-shifted with respect to $\lambda_l$).
- The second transmission minimum at **point 3** reflects the decrease of N. Again $\lambda_l$-$\lambda_r$ changes sign, this time from positive to negative values ($\lambda_r$ red-shifted with respect to $\lambda_l$). After point 3 the cavity is red-shifted and starts cooling down. In this region N decreases.
- After **point 4**, N stops decreasing and starts a slow increase.

The total time required to complete the {$\Delta T$ ,$N$} limit cycle (Fig. S4b) is $1/\nu_{SP}$, although the cycle is not drawn at a constant pace.
A necessary ingredient for the activation of the SP is the time delay of $\Delta T$ with respect to $N$, which can be seen in the temporal traces (see the zoomed panels of Fig. S4a) but is evident in the limit cycle (Fig. S4b). Indeed, the maximum values of $N$ and $\Delta T$ are achieved at different positions of the curve.

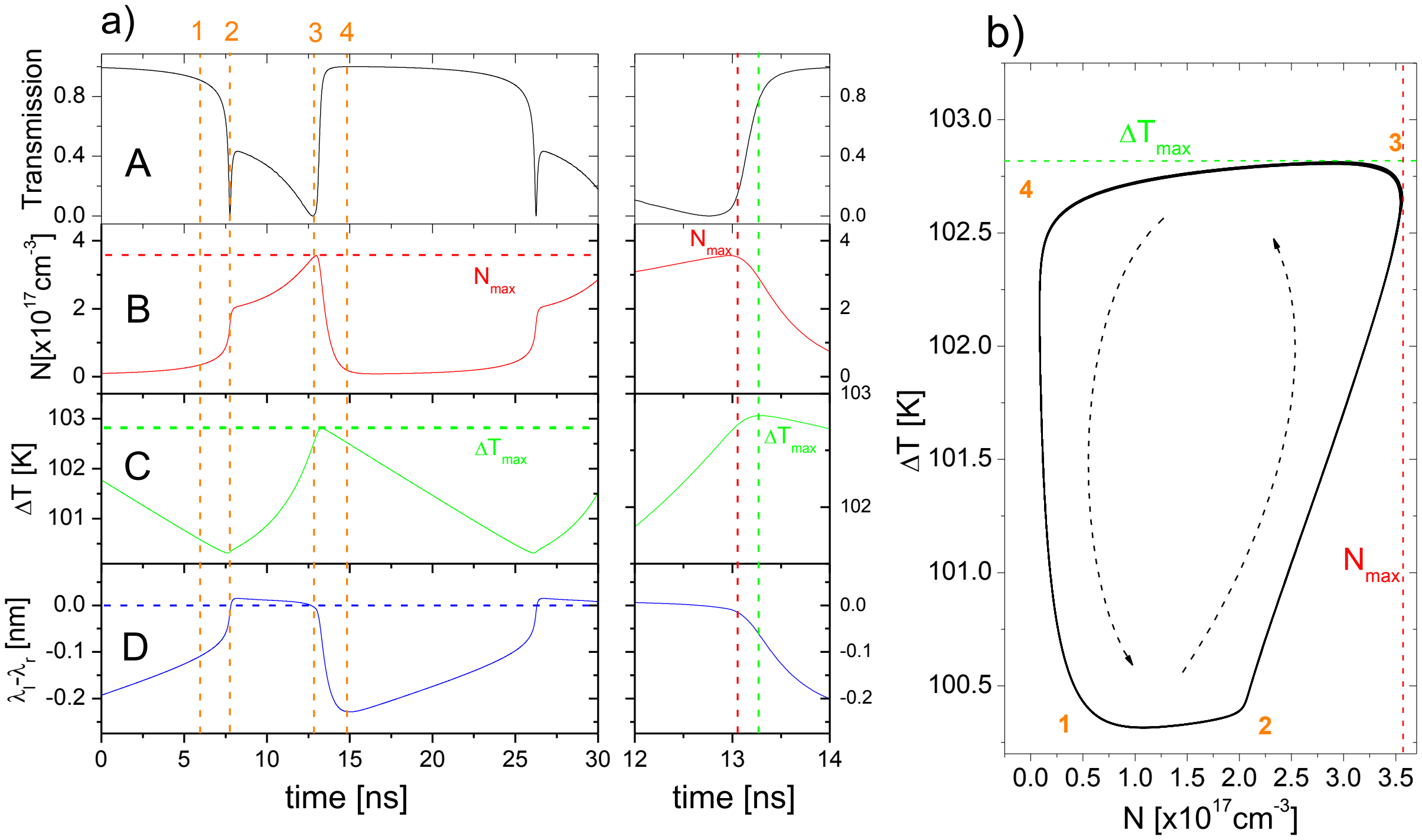

**Fig. S4. a)** Temporal behaviors of the transmitted optical signal, $N$, $\Delta T$ and $\lambda_l$-$\lambda_r$ (panels A, B, C and D respectively). Specific points in time are indicated with dashed vertical lines. The maximum values of $N$ and $\Delta T$ (panels B and C) and the condition $\lambda_l = \lambda_r$ (panel D) are indicated with dashed horizontal lines. A zoomed temporal region is represented in the right panels. **b)** $\{\Delta T ,N\}$ limit cycle. The rotation sense is indicated with black dashed arrows. The maximum values of N and ΔT are also indicated with red and green arrows, respectively.

### *S4.2. Self-pulsing frequency dependence with $\lambda_l$*

Since the $\Delta T$ decay rate depends on its absolute value, $\nu_{SP}$ increases with $\lambda_l$ as the cavity resonance is pushed to longer wavelengths. This is clearly observed when comparing, in the same time scale, the temporal traces of the SP for two well separated values of $\lambda_l$ (Fig. S5). As expected, the absolute value of the TO contribution (red curves of Fig. S5, note that when the temperature decreases the curves go up) decreases much faster for the longer $\lambda_l$. It is also observed that the duty cycle, i.e., the ratio between the time lapse in which TO and FCD are competing and the signal period, also increases with $\lambda_l$.

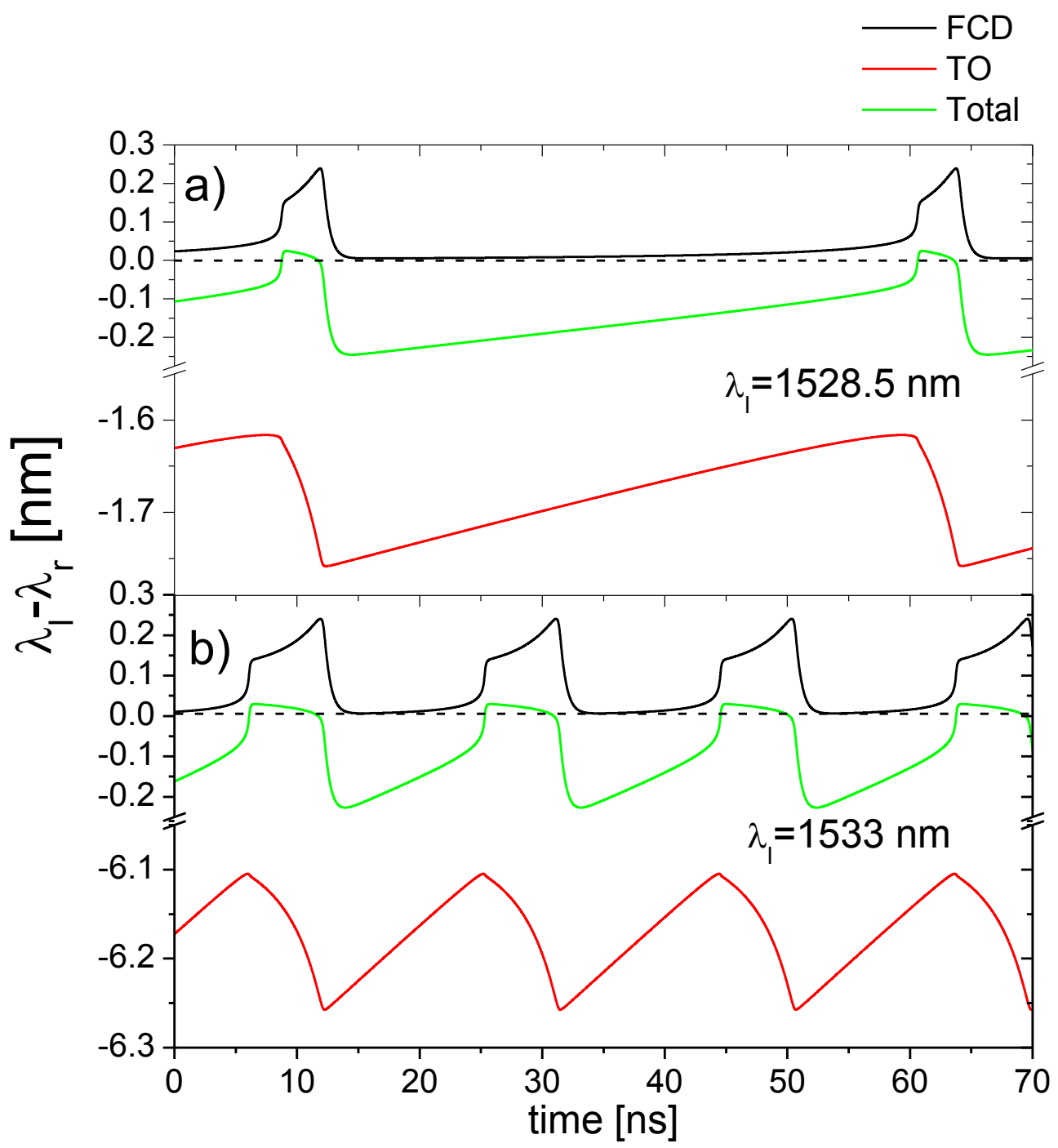

**Fig. S5.** Simulated temporal profiles of the FCD (black) and TO (red) contributions to the spectral shift of $\lambda_r$ for $\lambda_l$=1528.5 nm (panel **a**) and $\lambda_l$=1533 nm (panel **b**). The overall spectral shift is represented in green.

A five-fold enhancement of $\nu_{SP}$ from 10 to 50 MHz is shown on Fig. S6, where we represent the FFT of the intracavity optical force. Due to its anharmonicity it is composed by a comb of frequencies, in accordance with the experimental data.

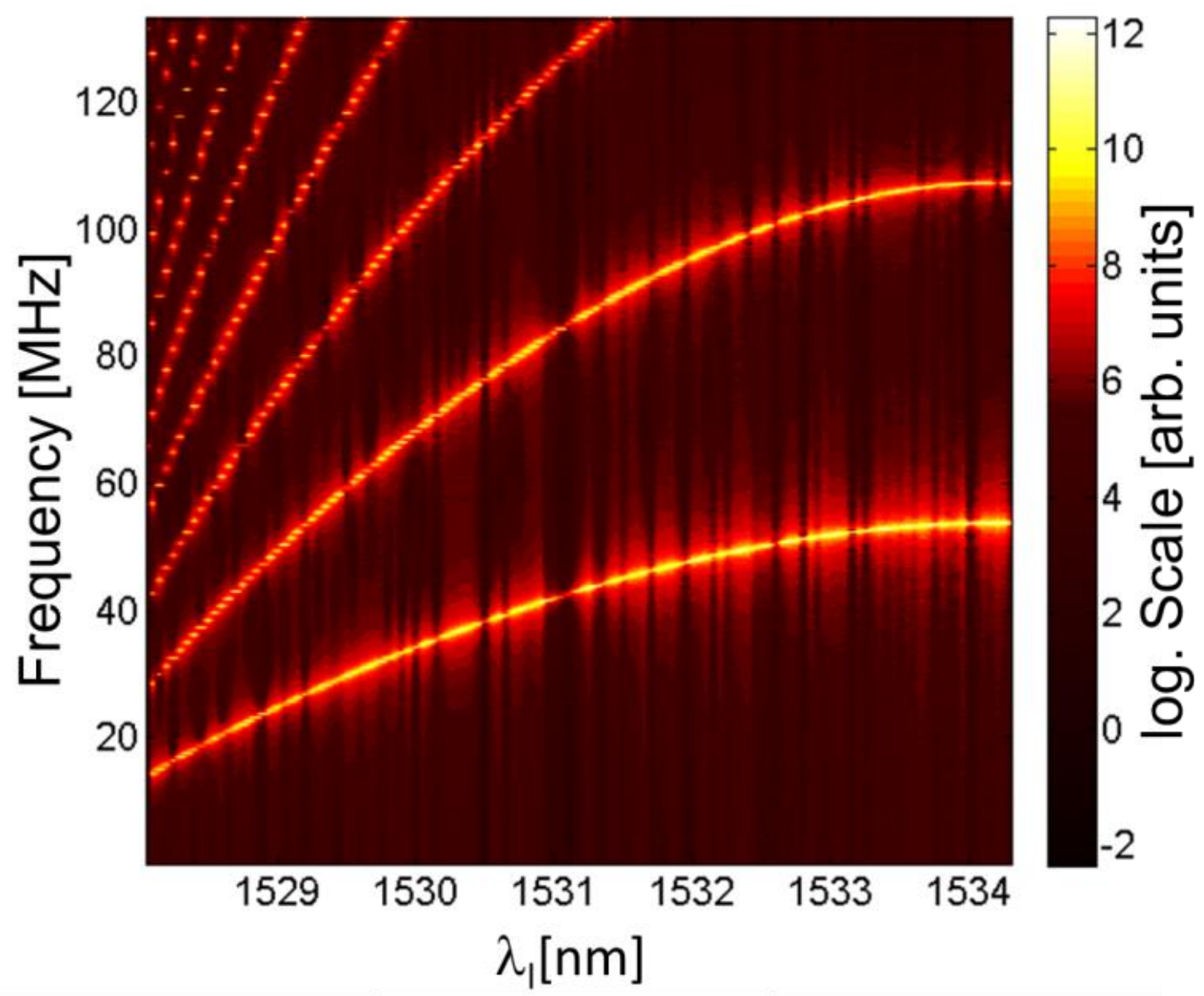

**Fig. S6.** Color contour plot of the FFT of the intracavity optical force (in log scale).

## *S5.* RF signal and temporal traces dependence with $\lambda_l$

### *S5.1 Coherent mechanical amplification using different harmonics of the force and limit cycles {ΔT, N, u}*

When introducing Eqs. (3) and (4) in the model (see main text), the FFT of the optical force as a function of $\lambda_l$ (Fig. S7a) presents several frequency-entrained (FE) regions at simple fractions of the frequency of the mechanical modes. In those regions the motion of the OM photonic crystal is coherently amplified by one of the harmonics of the force. FigS7b illustrates the former concept by showing the emergence of a peak at $\nu_m$ when one of the low harmonics of the force main peak at $\nu_{SP}$ is resonant with the mechanics. The stationary solutions for the mechanical displacement are almost pure harmonic signals with small components at the harmonics of $\nu_{SP}$.
In Fig S7 we highlight three situations: I and III cuts fall within FE regions with M=2 and M=1 respectively, II represents a frequency-unlocked region.
Limit cycles are obtained when representing any two variables of the set {*ΔT*, *N*, *u*} versus each other within regions I and III (black and green curves of Fig. S7c). In both previous cases the frequencies of the two oscillators are entrained/locked.
In region II the OM photonic crystal motion cannot be efficiently pumped by any of the harmonics of the force and is forced to oscillate non-resonantly, exhibiting low amplitude and a noisy {*N*, *u*} or {*ΔT*, *u*} limit cycles (red curve of Fig. S7c). A stable {*ΔT*, *N*} limit cycle is obtained also in this case.

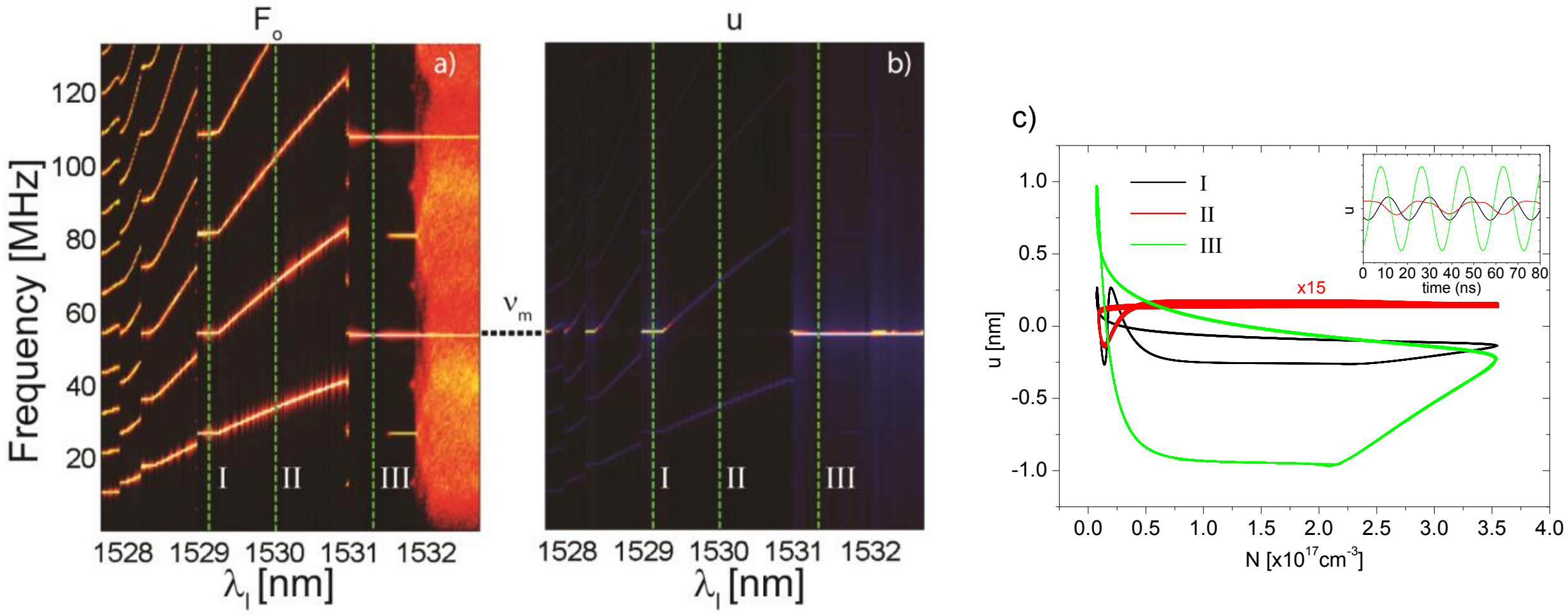

**Fig. S7. a)** and **b)** Simulated FFT (in log scale) of the optical force (panel a) and *u* (panel b). The vertical cuts correspond to a M=2 and M=1 frequency-locked state (I and III respectively) and to a frequency-unlocked state (II), **c)** {*N*, *u*} limit cycles on states I, II and III (black, red and green curves respectively). Note that, for the II case, *u* has been enhanced by a factor 15 for comparison. The simulations have been made only considering the 2nd in-plane odd flexural mode ($\nu_m$=54MHz) and its corresponding single-particle OM coupling rate, i.e., $g_{o,OM}/2\pi$=50 kHz.

In the experiment, mechanical modes at frequencies higher than 120 MHz could not be amplified coherently because the strength of the force decays with the harmonic number and cannot compensate the low $Q_m$ values measured at atmospheric conditions. Lower temperatures and pressures of the external environment would increase the $Q_m$ values, possibly enabling amplification in the GHz range, where the localized modes are present (*2*).

### *S5.2 Frequency and duty cycle dependence on $\lambda_l$ within frequency locked and unlocked regions.*

In Section 4.2, we showed that the $\Delta T$ decay rate increases with $\lambda_l$. Within a frequency-unlocked region, the former behavior leads to an increasing of $\nu_{SP}$. On the contrary, within a FE region, the $\Delta T$ decay rate increase is compensated by a monotonous increase of the duty cycle (temporal width of the asymmetric transmission minimum), so that $\nu_{SP}$ is constant. In Fig. S8a we show the experimental case of M=1, where $\nu_{SP}$ locks to $\nu_m$.
This feature is also reproduced by our model when the SP is coupled to the mechanics. In FigS8b we show the simulated temporal traces for two extreme points within the M=1 FE region.

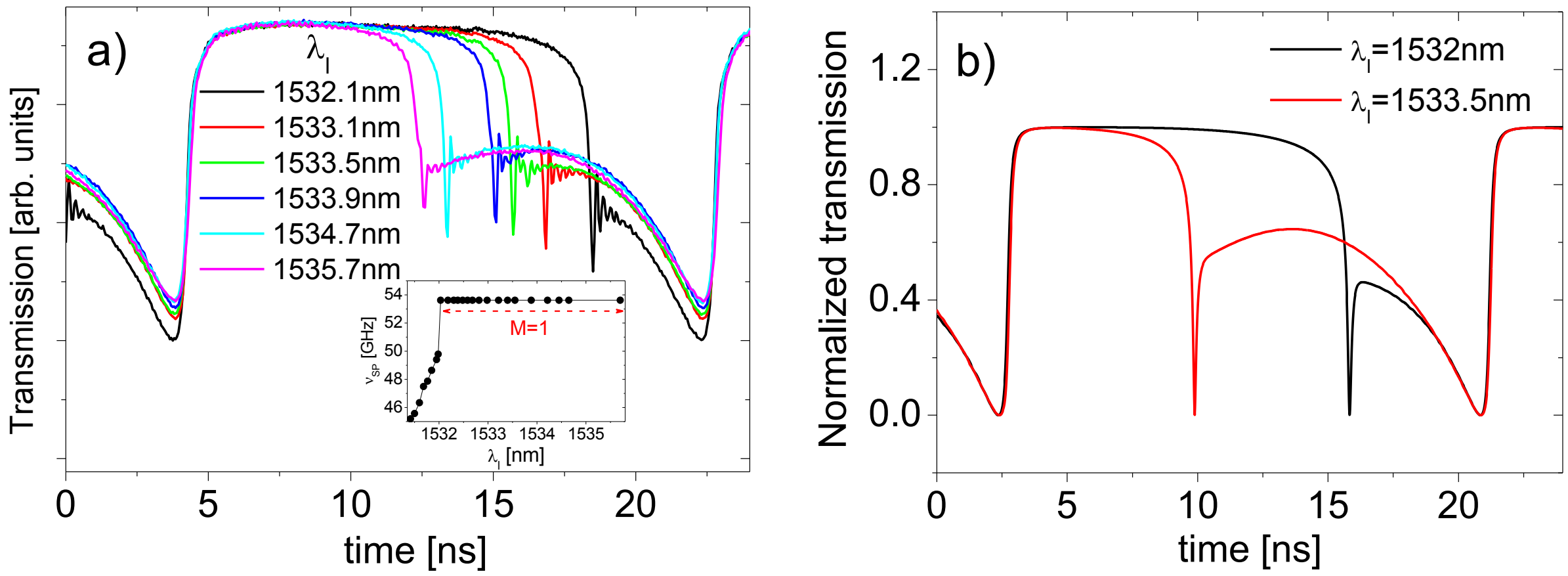

**Fig. S8. a)** Temporal traces for different values of $\lambda_l$ within the M=1 FE region. The inset shows the experimental Devil's staircase plot for reference **b)** Simulated transmission for two extreme values of $\lambda_l$ within the M=1 FE region. The simulations have been made only considering the $2^{nd}$ in-plane odd flexural mode ($\nu_m$=54MHz) and its corresponding single-particle OM coupling rate, i.e., $g_{o,OM}/2\pi$=50 kHz.

The RF spectra corresponding to extreme situations within the M=1 FE region (Fig. S9) show that the spectral position of the harmonics is obviously the same. However, there is a significant difference on their relative weights, which is consistent with the differences observed in the temporal signals of Fig. S8a.

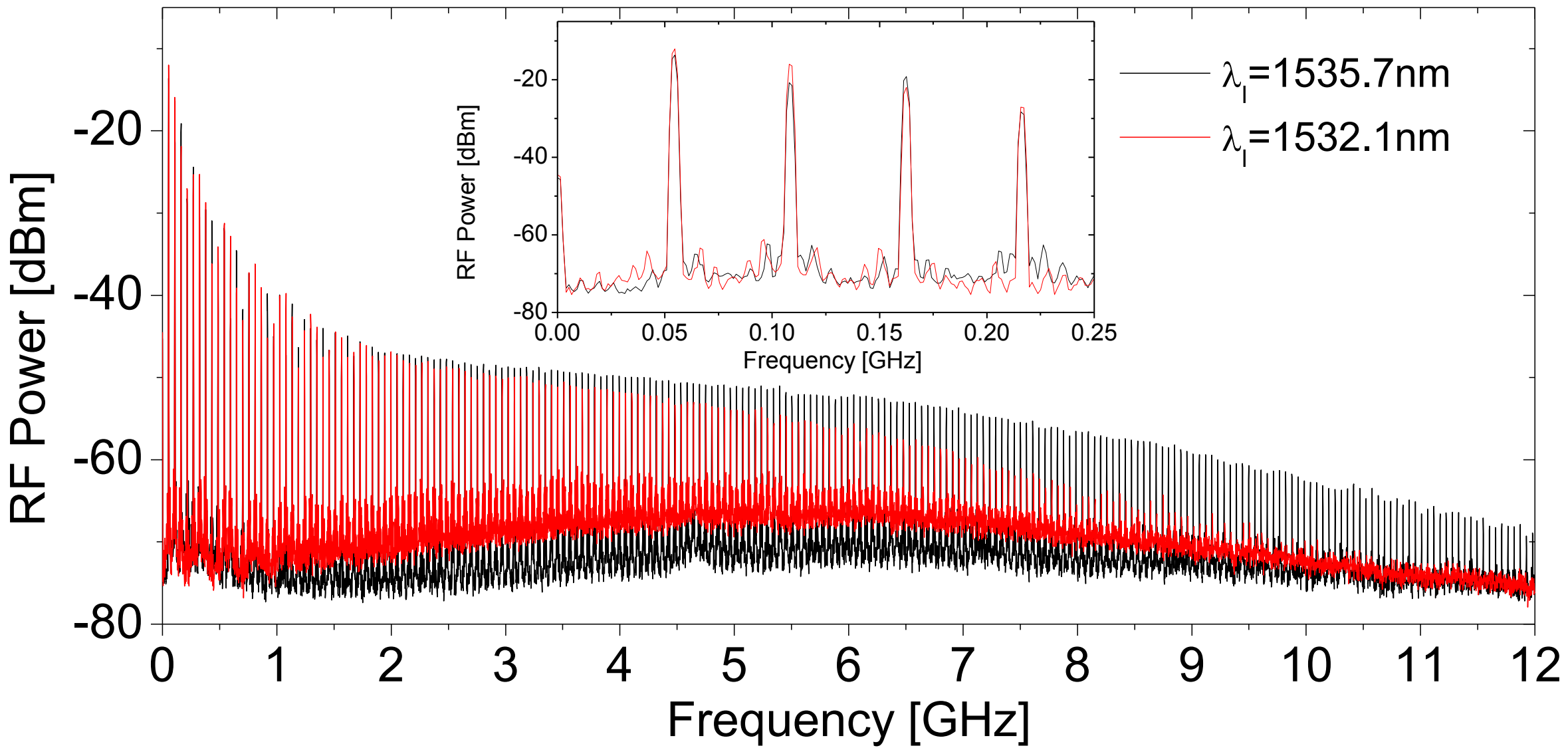

**Fig. S9.** RF signal obtained at the two extremes of the M=1 FE region.

The RF peaks in a frequency-unlocked region (Fig. S10) are inhomogenously broadened in frequency because of the integration time, i.e., the unlocked resonance central frequency oscillates fast in time (less than μs) (see Supplementary Videos S1 and S2). As a consequence, at frequencies above ~1GHz, there is an overlapping of harmonics of different frequencies, which results in a white signal.

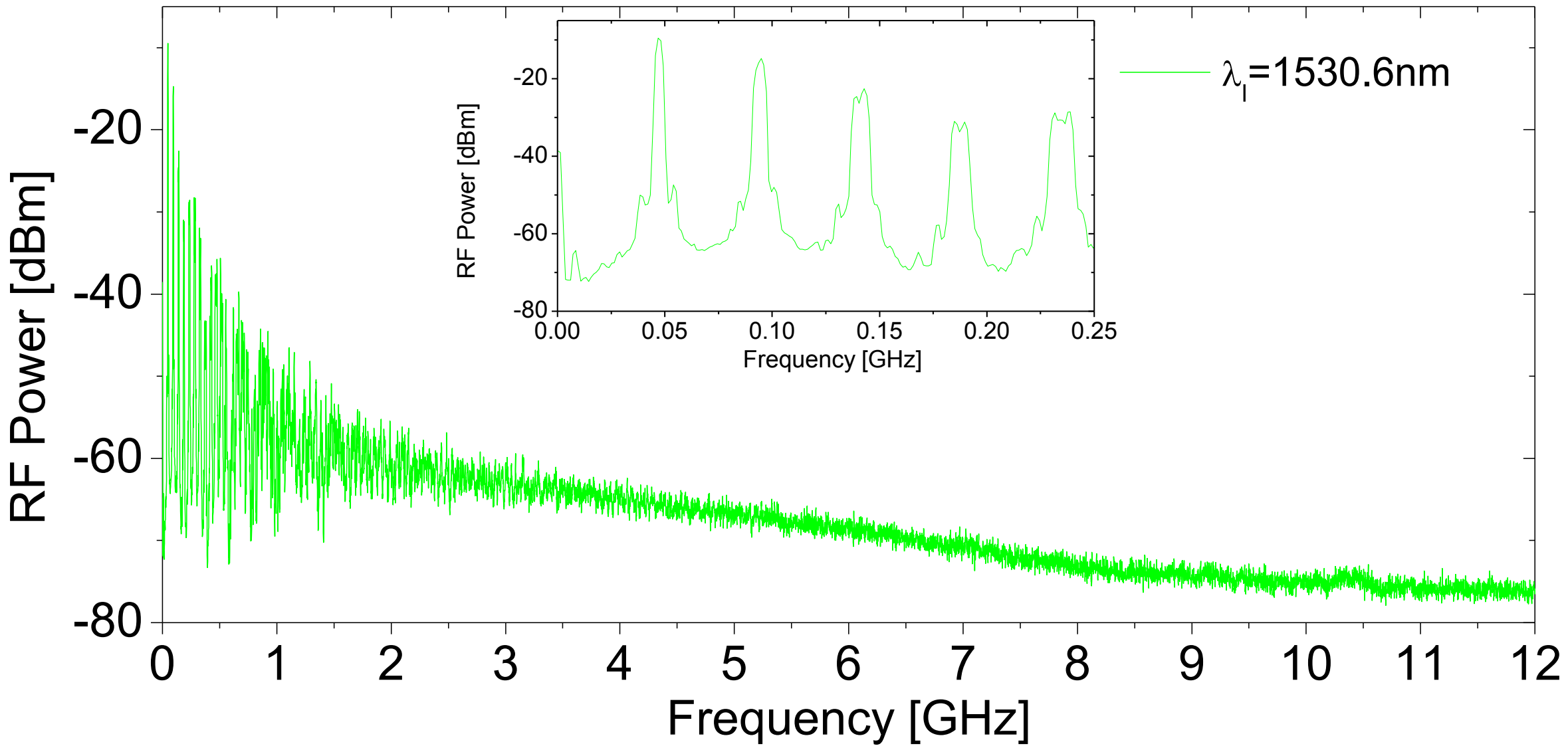

**Fig. S10.** RF signal obtained on a frequency-unlocked region.

### S6. Transduction of high frequency mechanical modes when SP is active.

The activation of the SP affects the transduced signal associated to localized mechanical modes in the GHz range. In Fig. S11 we show 2D plot of the RF spectrum of a mode at $\nu_m$=7.3 GHz and its dependence with $\lambda_l$. It is worth noting that this particular measurement has been obtained by exciting an optical mode different than the one used throughout the manuscript.
Above the threshold for SP (white dashed line) there is a strong degradation of the transduced signal of one order of magnitude. Moreover, the transduced peak becomes a central peak with equally spaced sidebands that spread apart when $\lambda_l$ is further increased. This feature is explained in terms of the activation of the SP. In that situation, the thermally activated motion of the localized mode weakly hypermodulates the anharmonic transmitted signal. Consequently, the RF signal associated to the localized mode is no longer a single peak placed at $\nu_m$ but a symmetric comb of peaks at $\nu_m \pm M\nu_{SP}$ (M=0, 1, 2, …). It is worth noting that, under the experimental conditions of Fig. S11, the SP is unlocked along the whole range covered by $\lambda_l$.

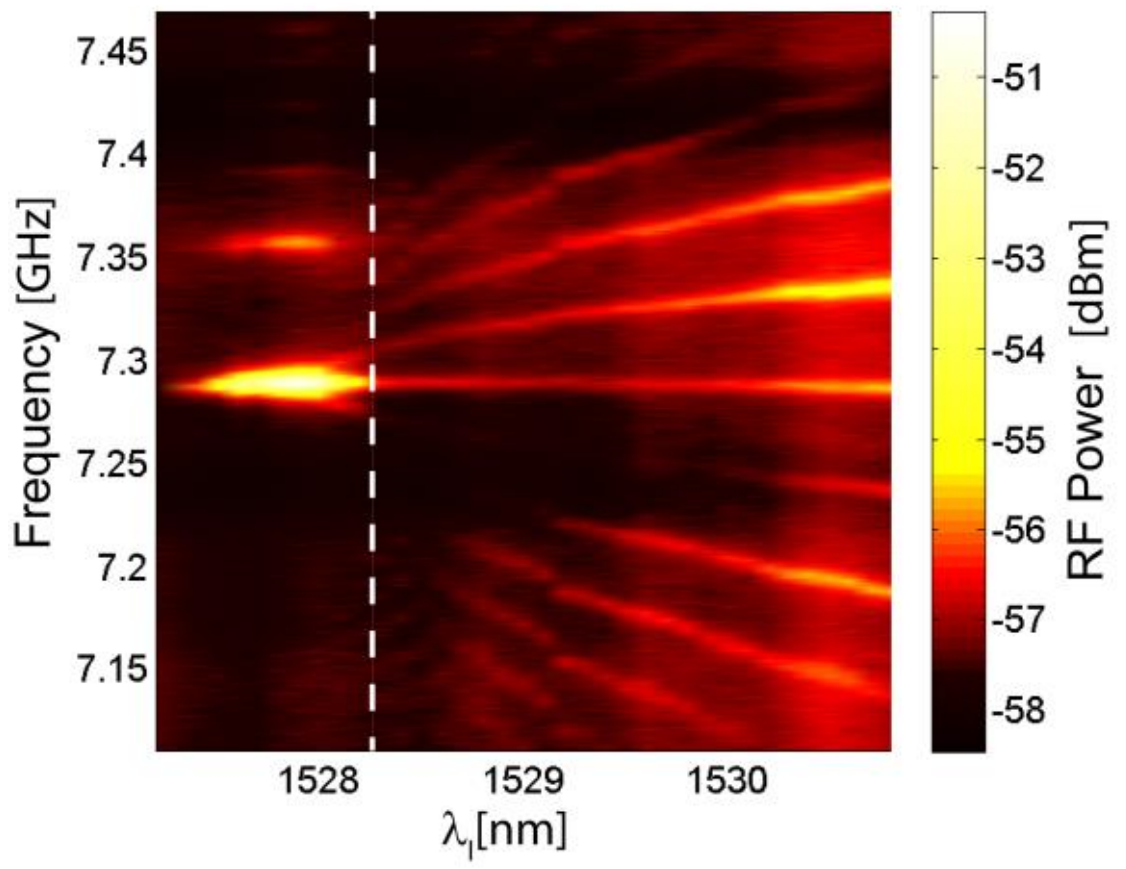

**Fig. S11.** Color contour plot of the RF power of a localized mechanical mode at $\nu_m$ = 7.3 GHz obtained for $P_l$=0.5 mW. The threshold for the SP activation is indicated with a white vertical line.

**Movies**

Video S1: RF signal as a function of laser wavelength with a frequency span of 200MHz

Video S2: RF signal as a function of laser wavelength with a frequency span of 6GHz

Video S3: Temporal signal as a function of laser wavelength